\documentclass[prd,preprintnumbers,twocolumn,superscriptaddress,nofootinbib]{revtex4-1}
\pdfoutput=1

\usepackage[utf8]{inputenc}
\usepackage[T1]{fontenc}
\usepackage{amsmath}
\usepackage{float}
\usepackage{graphicx}
\usepackage{comment}
\usepackage[dvipsnames]{xcolor}
\usepackage{xcolor}
\usepackage{color}
\usepackage{epsfig}
\usepackage{dcolumn}
\usepackage{hyperref}
\usepackage{bm}
\usepackage{amssymb}
\usepackage{slashed}
\usepackage{epstopdf}
\usepackage[normalem]{ulem}	
\usepackage{ae}
\pdfoutput=1

\makeatletter
\renewcommand{\maketag@@@}[1]{\hbox{\m@th\normalsize\normalfont#1}}%
\makeatother

\definecolor{red}{rgb}{1,0,0}
\definecolor{darkpink}{rgb}{0.8,0,0.4}

\DeclareRobustCommand{\Eq}[1]{Eq.~(\ref{#1})}
\DeclareRobustCommand{\Fig}[1]{Fig.~\ref{#1}}

\hypersetup{
colorlinks=true,
urlcolor=corlinks,
linkcolor=corlinks,
menucolor=cormenu,
citecolor=corlinks,
pdfborder= 0 0 0
}

\newcommand{\ie}{\text{i.e.}}
\newcommand{\eg}{\text{e.g.}}
\usepackage{tikz,pgfplots}
\usetikzlibrary{arrows,shapes,trees,calc,positioning,patterns}

\definecolor{corlinks}{RGB}{0,0,128}
\definecolor{cormenu}{RGB}{0,0,128}
\definecolor{corurl}{RGB}{0,0,128}

\newcommand{\ew}{\textnormal{\tiny EW}}

\newcommand{\MPl}{M_\text{Pl}}

\newcommand{\eq}{\mathrm{eq}}

\newcommand{\keV}{\,\text{keV}}
\newcommand{\MeV}{\,\text{MeV}}
\newcommand{\GeV}{\,\text{GeV}}

\newcommand{\s}{\,\text{s}}

\allowdisplaybreaks 

\begin{document}

\title{Relaxion  Dark Matter}

\author{Nayara Fonseca}
\email{nayara.fonseca@desy.de}
\affiliation{DESY, Notkestrasse 85, 22607 Hamburg, Germany}
\author{Enrico Morgante}
\email{enrico.morgante@desy.de}
\affiliation{DESY, Notkestrasse 85, 22607 Hamburg, Germany}
%


\begin{abstract}

We highlight a new connection between the Standard Model hierarchy problem and the dark matter sector.  The key piece  is the relaxion field, which besides scanning the Higgs mass and setting the electroweak scale, also constitutes the observed dark matter abundance  of the universe.  The relaxation mechanism is realized during inflation, and the necessary friction is provided by particle production. Using this framework we show that the relaxion is a phenomenologically viable dark matter candidate  in the  keV mass range.
\end{abstract}

\preprint{DESY 18-161}
\maketitle


\section{Introduction}

Despite the impressive effort of the community, the  non-gravitational nature of dark matter (DM) is still unknown.  In light of  no  definitive evidence of new physics at the TeV scale and strong exclusion limits from direct detection experiments, to go beyond the WIMP (weakly interacting massive particle) paradigm became crucial.   In this work we propose another option which can closely connect the Higgs naturalness problem with the DM sector. The link is the relaxion field.

The cosmological relaxation of the electroweak scale is a recent proposal to address the Standard Model (SM) hierarchy problem making use of  the relaxion,  an axion-like field which scans the Higgs mass parameter during its cosmological evolution   \cite{Graham:2015cka}.
The main idea is very simple. The relaxion (which we will call $\phi$ for the rest of the paper) rolls down a linear potential $g\Lambda^3\phi$, where $g$ is a small coupling that parametrizes the explicit breaking of the shift-symmetry of $\phi$, which we assume it is a pseudo-Nambu-Goldstone-boson. The field $\phi$ also couples to the Higgs doublet through a term $\sim g' \Lambda \phi h^2$, where  $g'\ll 1$  and $\Lambda$ is assumed to be the cut-off of the theory, which also controls the Higgs mass term $\Lambda^2 h^2$.%
\footnote{To simplify the notation, we will treat the Higgs as a single real scalar field, whose evolution is purely classical, and all fluctuations will be neglected.}
During its evolution, the relaxion provides an effective mass term for the Higgs that varies with time, until the evolution is stopped when the Higgs mass squared is negative and has the size of the electroweak scale ($v_\ew^2$).

The relaxion evolution stops due to a back-reaction mechanism  which turns on when the Higgs vacuum expectation value (VEV) is at the electroweak scale.
For example, in Ref.~\cite{Graham:2015cka} a term of the form $\Lambda_b^4\cos(\phi/f)$ was added to the potential, where $\Lambda_b^4$ depends on the Higgs VEV $h$. As EW symmetry breaks, the Higgs VEV and thus $\Lambda_b$ grow until the velocity is not large enough to overcome the barriers provided by the cosine potential. After that point the relaxion is trapped and the EW scale is determined by the final value of $\phi$.
An alternative realization, which we will consider in this paper, was introduced in Ref.~\cite{Hook:2016mqo}. In this case, the barriers $\Lambda_b^4\cos(\phi/f)$ are constant, and the field's kinetic energy is large enough to overcome them. By assumption, the EW symmetry is broken early on, and all the SM particles are initially very heavy. When the relaxion approaches the critical point at which the Higgs VEV is zero, all SM particles become light. At this point, the relaxion stops due to the production of SM gauge bosons, due to a coupling that we will discuss below in Sec.~\ref{sec:relaxion pp}. The EW scale can be related to the parameters of the model, with $v_\ew\ll\Lambda$ in a technically natural way.

This paradigm shift fits in the interface between particle physics and  early universe cosmology and gave rise to a varied literature, including
studies on the model building  challenges
\cite{Espinosa:2015eda, Gupta:2015uea, Abel:2015rkm, Choi:2015aem, Ibanez:2015fcv, Hebecker:2015zss, McAllister:2016vzi, Fonseca:2016eoo, Nelson:2017cfv, Gupta:2018wif},
concerns about the inflationary and reheating sectors
\cite{Patil:2015oxa, Jaeckel:2015txa, Marzola:2015dia, DiChiara:2015euo,Tangarife:2017rgl, Choi:2016kke},
alternatives to inflation
\cite{Hardy:2015laa, Hook:2016mqo, Fonseca:2018xzp},
UV completions
\cite{Batell:2015fma, Evans:2016htp, Evans:2017bjs, Fonseca:2017crh},
developments on  the model building front
\cite{Batell:2017kho, Antipin:2015jia,Agugliaro:2016clv, Lalak:2016mbv, Matsedonskyi:2015xta, Davidi:2017gir, Huang:2016dhp, Matsedonskyi:2017rkq, Davidi:2018sii, Wang:2018ddr, Gupta:2019ueh, Ibe:2019udh},
baryogenesis
\cite{Son:2018avk, Abel:2018fqg},
experimental signatures
\cite{Kobayashi:2016bue, Choi:2016luu, Flacke:2016szy, Beauchesne:2017ukw, Frugiuele:2018coc},
and cosmological implications
\cite{Banerjee:2018xmn, Banerjee:2019epw}.

The cosmological consequences of the relaxion scenario have not been fully explored yet. This is of the maximum importance, as it would help both in constraining the properties of the relaxion field and in pointing out possible observable signatures of its existence. In this note, we address the question of whether the relic population of relaxion particles can constitute the current DM density.
In previous literature, the answer was negative.%
\footnote{After this paper appeared, Ref.~\cite{Banerjee:2018xmn} pointed out the possibility of obtaining relaxion DM through the coherent oscillations of the relaxion field after reheating, using the model of Ref.~\cite{Graham:2015cka}. If the reheating temperature is larger than the scale $\Lambda_b$, the relaxion is displaced from its minimum and its oscillations carry a sizeable energy.}
The constructions discussed in Ref.~\cite{Graham:2015cka, Espinosa:2015eda,Flacke:2016szy}  use a Higgs-dependent barrier  to stop the field evolution, which happens during inflation. In these models,  the relaxion  abundance is negligible~\cite{Flacke:2016szy}. A second field, which scans the barriers' amplitude in Ref.~\cite{Espinosa:2015eda}, can instead have a sizable misalignment yield.
Oppositely, if relaxation proceeds after inflation, and the source of friction is the tachyonic production of gauge bosons, the relaxion is overproduced ~\cite{Fonseca:2018xzp}, and one has to  impose that its lifetime is short enough to dilute this abundance before nucleosynthesis.

Here we assume that relaxation happens during inflation and the relaxion stopping mechanism is provided by particle production \cite{Hook:2016mqo}.  This construction does not require new physics close to the TeV scale, and, in a portion of the parameter space, it can be realized without a large number of e-folds  or super-Planckian field excursions.
As we will detail below, in this scenario the relaxion particles, produced after reheating by scatterings in the SM plasma, can account for the observed DM density.

The paper is structured as follows. In Sec.~\ref{sec:relaxion pp} we introduce the model and discuss in details the mechanism to generate a small EW scale. In Sec.~\ref{sec:parameter space} we discuss the conditions that need to be applied on the parameters of the model. Section~\ref{sec:DM} discusses the production of the relaxion DM population in the early universe, whose properties are discussed in Sec.~\ref{sec:results}. Finally, we draw our conclusions in Sec.~\ref{sec:conclusions}.

%
\section{Relaxation with particle production}\label{sec:relaxion pp}

We now introduce the relaxion model that we will consider throughout this paper, which was first introduced in Ref.~\cite{Hook:2016mqo}. We will assume that relaxation takes place during inflation, and refer the reader to Ref.~\cite{Fonseca:2018xzp} for an analysis of the case in which relaxation happens after inflation.
The  Lagrangian is \cite{Hook:2016mqo}:
\begin{eqnarray} \nonumber
\mathcal{L} &\supset&  \frac{1}{2} \left(\Lambda^2  -g'\Lambda \phi  \right)h^2 + g \Lambda^3 \phi - \frac{\lambda}{4} h^4  - \Lambda_b^4\cos\left(\frac{\phi}{f'}\right)  \\  &-&\! \frac{\phi}{4\mathcal{F}} \left(g_2^2 W^a_{\mu\nu}\widetilde{W}^{a\,\mu\nu}\! - \!g_1^2 B_{\mu\nu}\widetilde{B}^{\mu\nu} \right),
\label{eq:relaxion_Lagrangian}
\end{eqnarray}
where  $\phi$ is  an axion-like field with decay constant $f'$, $h$ is the Higgs field,  $\Lambda$  is the cutoff of the theory,  the dimensionless parameters $g$ and $g'$ are assumed to be spurions that explicitly break the axion shift symmetry and $\lambda$ is the Higgs  quartic coupling. The scale $\Lambda_b$ is related to the confinement scale $\Lambda_c$ of some non-abelian gauge group, at which the $\phi$ cosine potential is generated as $\Lambda_b^4 \sim \epsilon \Lambda_c^4$, where in general $\epsilon\ll1$. For example, if the relaxion is the QCD axion, $\epsilon \approx m_q/\Lambda_\text{QCD}$, where $m_q$ is the mass scale of the up- and down-quark. In the following we will not specify any further the mechanism responsible for the generation of these barriers. The effective scale $\mathcal F$ controls the interaction of the relaxion with the SM gauge bosons. We assume that the relaxation dynamics takes place in the broken phase so that the Higgs mass parameter, $\mu_h^2(\phi)\equiv (-\Lambda^2 + g'\Lambda\,\phi)$, is large and negative when the scanning process starts, $\mu_h^2(\phi_{\textrm{ini}})\sim -\Lambda^2  $.
The potential of \Eq{eq:relaxion_Lagrangian} also induces a mixing of the relaxion with the Higgs~\cite{Flacke:2016szy}. After relaxation ends and the relaxion stops in one of the minima of its potential, the mixing angle is
\begin{equation}\label{eq:mixing}
\theta\approx \frac{g'v_\ew\Lambda}{\left[(m_h^2-m_\phi^2)^2+4g'^2v_\ew^2\Lambda^2\right]^{1/2}} \,.
\end{equation}
The last term in \Eq{eq:relaxion_Lagrangian} is  responsible to slow down the relaxion once the particle production is triggered. $B$ and $W$ are the SM gauge bosons with  $g_1$ and $g_2$ being the  corresponding $\textrm{U}(1)$ and $\textrm{SU}(2)$ gauge couplings.
When expanded in the mass eigenstates this term reads
\begin{align}\label{eq:lagbroken}
 - \frac{\phi}{\mathcal{F}} \epsilon^{\mu\nu\rho\sigma} & \Big(
2 g_2^2 \partial_\mu W^-_\nu \partial_\rho W^+_\sigma +
(g_2^2-g_1^2) \partial_\mu Z_\nu \partial_\rho Z_\sigma \nonumber \\
& - 2g_1 g_2 \partial_\mu Z_\nu \partial_\rho A_\sigma
\Big)\,.
\end{align}
In what follows  we will only consider the tachyonic instability from the $Z\widetilde{Z}$ term and  absorb the  gauge coupling in the definition of the corresponding field such that $1/f = (g_2^2 -g_1^2)/\mathcal{F}$.
The contribution from the $WW$ term is expected to be suppressed due to the self-interactions of the $W$, which induce an effective mass making  particle creation inefficient. Tachyonic production of photons, as we will see, is suppressed in this model, thus we expect the $ZA$ term in \Eq{eq:lagbroken} to be subdominant with respect to the $ZZ$ one.

As it is well-known in the context of axion inflation (see, e.g., Ref.~\cite{Domcke:2018eki} and references therein for a recent account), a coupling of the form $\phi F\widetilde F$ with $\dot\phi\neq0$ leads to the exponential production of gauge bosons. In the case at hands, this production is suppressed by the mass of the $Z$, which is initially large. As we will see below, particle production starts only when $m_Z\ll\Lambda$. At this point it should be clear that a coupling to photons $\phi F\widetilde F$ must be suppressed, because it would lead to exponential photon production during the relaxion's evolution, thus slowing down the field  independently of the Higgs VEV and spoiling the mechanism. The non-generic coupling structure of~\Eq{eq:lagbroken} is designed for this purpose. This can, for example, descend from a left-right symmetric UV completion~\cite{Craig:2018kne}. Also, it appears in non-minimal composite Higgs models with coset $\mathrm{SO}(6)/\mathrm{SO}(5)$~\cite{Gripaios:2009pe, Cacciapaglia:2014uja, Gripaios:2016mmi, Chala:2017sjk, Molinaro:2017mwb}, even an attempt of embedding the relaxion mechanism in such a construction is still missing in the literature.

The  interaction of \Eq{eq:lagbroken} generates  a coupling to  SM fermions (at one loop) and to  photons (at one and two loops) \cite{Bauer:2017ris, Craig:2018kne}:
\begin{equation}\label{eq:relaxion coupling fermions photons}
\frac{\partial_\mu \phi}{f_F} (\bar{\psi} \gamma^\mu \gamma_5 \psi)
\quad\text{and}\quad
\frac{\phi}{4f_\gamma} F \widetilde F
\end{equation}
where
\begin{equation}\label{eq:f_fermions}
\frac{1}{f_F} = \frac{3 \alpha_{\textrm{em}}^2}{4 \mathcal{F}}\left[  \frac{Y_{F_L}^2 +Y_{F_R}^2}{\cos^4\theta_W} -\frac{3}{4 \sin^4\theta_W}
\right]\log\frac{\Lambda^2}{m_W^2},
\end{equation}
and
\begin{equation} \label{eq:f_gamma}
\frac{1}{f_\gamma} =  \frac{2 \alpha_{\rm em}}{ \pi \sin^2\theta_W \mathcal{F}} B_2\left(x_W\right)+\sum_F \frac{N_c^F Q_F^2}{2 \pi^2 f_F}   B_1\left( x_F\right),
\end{equation}
where $N_c^F$ is  the color factor,  $Q_F$ is the electric charge of the fermion $F$ with mass $m_F$, and $x_i \equiv 4 m_i^2/m_\phi^2$. The functions $B_{1,2}$ are:
\begin{eqnarray}
\begin{aligned}
    B_1(x) = 1 - x [f(x)]^2, ~    B_2(x) = 1 - (x - 1) [f(x)]^2\!, ~~
  \end{aligned}
\\ f(x)=\begin{cases}
    \arcsin \frac{1}{\sqrt{x}},   &x \geq 1\\
    \frac{\pi}{2} + \frac{i}{2} \log \frac{1 + \sqrt{1-x}}{1 - \sqrt{1-x}} ,  &x < 1.~~~
  \end{cases}
\end{eqnarray}
When the relaxion is light, which will turn out to be the case of interest for our DM scenario, these functions scale like $B_1(x_F)\rightarrow -m_\phi^2/(12 m_F^2)$ and $B_2(x_W)\rightarrow m_\phi^2/(6 m_W^2)$ as $m^2_\phi\rightarrow 0$. This implies, for instance, that when the relaxion is lighter than the electron mass, the induced coupling to photons originated from the coupling in \Eq{eq:relaxion_Lagrangian} is suppressed.

Let us now describe how the relaxion mechanism works in this model. The $\phi$ field rolls down its potential  until it reaches the critical point when there is an exponential production of gauge bosons, which  makes the relaxion slow down due to the transfer of its energy to the gauge bosons.
This back-reaction mechanism becomes apparent once we examine the equations of motion for $\phi$ and $Z$: 
\begin{align}
\ddot\phi - g\Lambda^3 + g'\Lambda h^2 + \frac{\Lambda_b^4}{f'}\sin\frac{\phi}{f'} + \frac{1}{4f}\langle Z\widetilde{Z} \rangle &= 0, \label{eq:phieomvector} \\
\ddot Z_\pm +(k^2 + \left(m(h)\right)^2 \mp k\frac{\dot\phi}{f})Z_\pm &= 0, \label{eq:Veomvector}
\end{align}
where $m(h)=\sqrt{g_1^2+g_2^2}h/2$,  $\langle Z\widetilde{Z} \rangle$ is the expectation value of the quantum operator and $Z_\pm $ refers to the two transverse polarizations of $Z_\mu$.%
\footnote{We neglect the longitudinal mode $Z_L$ as it does not have a tachyonic instability.}
In terms of the mode functions $Z_\pm$, $\langle Z\widetilde{Z} \rangle$ can be written as
\begin{equation}
\langle Z\widetilde{Z} \rangle = \int\frac{d^3k}{(2\pi)^3}\left(|Z_+|^2-|Z_-|^2\right) \,.
\end{equation}
Assuming $\dot\phi>0$, $Z_+$ has a  tachyonic growing mode when $\omega_{k,+}^2 \equiv k^2 + \left(m(h)\right)^2 - k\frac{\dot{\phi}}{f} <0$.  The first mode that becomes tachyonic is the one for which $\omega_{k,+}$ is minimum, $k_c=\dot{\phi}/(2f)$,
 thus $Z_+$ grows exponentially for
\begin{equation} \label{eq:tachyon}
\dot{\phi} > 2 f\, m(h).
\end{equation}
The growth of the mode function $Z_+$ continues until $\langle Z\widetilde Z\rangle$ becomes the dominant term in the equation of motion of $\phi$, \Eq{eq:phieomvector}. After this point, $\ddot\phi\approx -\langle Z\widetilde Z\rangle/(4f)<0$ and the relaxion velocity decreases.
After $\phi$ has slowed down, the constant cosine potential acting as a barrier can make the relaxion evolution stop.
An example of such evolution was computed numerically in Ref.~\cite{Fonseca:2018xzp}.

The $Z$ mass depends on the Higgs field, implying that condition (\ref{eq:tachyon}) is not satisfied when the Higgs field value is large and particle production is ineffective. 
To obtain the correct value of the electroweak scale, the back-reaction should be triggered when $m(h) = m_Z \approx 90\GeV$, \ie, ~for
\begin{equation}\label{eq:stopping condition}
\dot{\phi} \sim 2f  m_Z \,.
\end{equation}
In the following we will use this equation to write $f$ as a function of the model's parameters.
The back-reaction must turn on when $\phi$ is close to the critical value $\Lambda/g'$ that cancels the Higgs mass term in \Eq{eq:relaxion_Lagrangian}, generating a parametric hierarchy between the cutoff  $\Lambda$ and the electroweak scale.
This happens if the classical Higgs field $h$ follows closely the minimum of its potential, as we will detail in the next section.

After the tachyonic growth starts, the Higgs and gauge fields undergo a complicate dynamics. First of all, we expect the system to quickly thermalize as soon as the energy density of the gauge fields becomes larger than the EW scale \cite{Hook:2016mqo} (see also Ref.\,\cite{Fonseca:2018xzp}). The interaction rate can be estimated as
\begin{equation}
\Gamma_\text{int} \sim n \langle\sigma v\rangle \sim \frac{\rho}{v_\ew} \frac{1}{v_\ew^2} \,,
\end{equation}
with $\rho \sim \dot\phi^2>v_\ew$, thus $\Gamma_\text{int} > v_\ew$,
which should be compared with the tachyonic growth rate which is $\mathcal{O}(v_\ew)$. Notice that, in the following, we will  impose that the  timescale for particle production is much shorter than the Hubble time.

Finite density effects affect both the Higgs and the gauge fields' evolution. A positive mass term for the Higgs is generated in the thermal bath, temporarily restoring the electroweak symmetry.
The field $h$ (and thus the mass of the gauge bosons) rolls to zero, making the tachyonic growth faster.

On the other hand, the presence of the thermal plasma affects the dispersion relation of the gauge bosons, which is modified into
\begin{equation}\label{eq:dispersion relation}
\omega^2_{k,+} = k^2 + m_V^2 - k\frac{\dot\phi}{f}
+ \Pi[\omega,k]
\end{equation}
where, in a hard thermal loop (\ie, high temperature) limit \cite{Bellac:2011kqa},
\begin{equation}\label{eq:dispersion relation 2}
\Pi[\omega,k] = m_D^2\frac{\omega}{k}\left(\frac{\omega}{k}+\frac{1}{2}\left(1-\frac{\omega^2}{k^2}\right)\log\frac{\omega+k}{\omega-k} \right)\,.
\end{equation}
Here $m_D^2 = g_\ew^2 T^2/6$ is the Debye mass of the bosons in the plasma. The factor $g_\ew$ is obtained by taking into account all the SM fermions and their respective hypercharges, as $g_\ew \approx (32/9) \sin^2\theta_W g_1^2 \approx 0.2$, where $g_1$  is the SM hypercharge coupling and the Weinberg angle $\sin^2\theta_W$ projects the $Z$ onto its abelian component.

The function $\Pi[\omega,k]$ is positive for imaginary frequency $\omega=i\Omega$, thus damping the tachyonic instability.
Expanding \Eq{eq:dispersion relation 2} for $\Omega / k \to 0$ we obtain
\begin{equation}
\Pi[\Omega,k] \approx \frac{\pi}{2}\frac{|\Omega|}{k}m_D^2\,.
\end{equation}
$\Omega$ is thus maximized for $k = 2\dot\phi/ (3 f)$, with
\begin{equation}\label{eq:Omega thermal}
\Omega_\textrm{max} \approx \frac{8}{27\pi}\frac{\dot \phi^3}{f^3 m_D^2} = \frac{16}{9\pi g_\ew^2} \frac{\dot \phi^3}{T^2 f^3}\,.
\end{equation}
From \Eq{eq:Omega thermal} we can estimate the typical timescale for the exponential growth as
\begin{equation}
\Delta t_{\rm{pp}} \sim \frac{9\pi g_\ew^2}{16} \frac{T^2 f^3}{\dot \phi^3} \,.
\end{equation}
The temperature $T$ is obtained  by assuming that the relaxion kinetic energy is transferred to radiation:
\begin{equation}
\frac{1}{2}\dot\phi^2 \sim \frac{1}{2}\left(\frac{V_\phi'}{3H_I}\right)^2 \sim \frac{\pi^2}{30}g_* T^4 \,.
\end{equation}
To summarize, we expect that particle production leads to the production of a thermal bath of SM particles, which temporarily restores EW symmetry and, at the same time, reduces the particle production rate. After the relaxion field stops, the temperature is rapidly erased by cosmic expansion, and the Higgs relaxes to its zero temperature VEV which is now fixed by the value of the relaxion field. Equation~(\ref{eq:stopping condition}) ensures that the final Higgs VEV is the measured one.

\section{Parameter space}\label{sec:parameter space}

The parameter space is characterized by six parameters: the cut-off $\Lambda$, the couplings $g$ and $g'$, the barriers' height $\Lambda_b$, the decay constant $f'$, and the Hubble constant during inflation $H_I$.
The scale $f$ can be fixed in terms of the other parameters  using \Eq{eq:stopping condition} $f = \dot\phi/(2 m_Z)$ and the value of the slow-roll velocity $\dot\phi= g\Lambda^3/(3 H_I)$, yielding
\begin{equation}\label{eq:f}
f = \frac{g \Lambda^3}{6 H_I m_Z} \,.
\end{equation}
The couplings $g,g'$ must satisfy $g>g'/(4\pi)^2$, otherwise a linear term $g'\Lambda^3\phi / (4\pi)^2$, generated through a Higgs loop, would dominate over the term $g\Lambda^3\phi$ in the relaxion potential.
Since the two spurions may be generated in a similar manner in the UV theory, it would be reasonable to assume $g=g'$. Still, as we will show below, it is convenient to relax this assumption, and therefore in the following we will consider the benchmarks $g/g' = 1, 10^3, 10^6$.

Independently of the relaxion being the DM, there are a number of conditions that the model must satisfy to actually solve the hierarchy problem.
First, the relaxion should not affect the inflationary dynamics, implying that the relaxion potential is subdominant compared to the inflaton one. This  gives a lower bound on the inflation scale $H_I$:
\begin{equation}
\label{eq:Hubblelarge}
V_\phi \sim \Lambda^4 \lesssim H_I^2 \MPl^2\,.
\end{equation}
In addition, the assumption that $\phi$  evolves classically is valid only if the classical evolution dominates over the quantum fluctuations during inflation. Therefore we impose that, over a Hubble time, $(\delta \phi)_{\rm class} \gtrsim (\delta \phi)_{\rm quant}$ with $(\delta \phi)_{\rm class} \sim V'_\phi/(3 H_I^2)$ and $(\delta \phi)_{\rm quant} \sim H_I/(2 \pi)$. This gives us an upper bound on the inflation scale,
\begin{equation} \label{eq:Hubblesmall}
 H_I \lesssim \left(\frac{ 2 \pi}{3} \right)^{1/3}\left(g \Lambda^3\right)^{1/3},
\end{equation}
where we used that $V'_{\phi} \sim g \Lambda^3$.

Furthermore, inflation should  last long enough such that the relaxion has time to scan the Higgs mass parameter. 
 The minimal number of e-folds  which is  required to scan a field range $\Delta \phi \sim \Lambda/g'$, is given by
\begin{equation} \label{eq:Ne}
\mathcal{N}_e  \sim (\delta \phi_{\textrm{class}})^{-1}\frac{\Lambda}{g'}
\sim \frac{3 H_I^2}{g' g \Lambda^2}
\sim \frac{g\, \Lambda^4}{12\, g' m_Z^2 f^2} \,,
\end{equation}
where in the last step we used that the slow roll velocity is $\dot{\phi} \sim 2m_Zf$.

We also need  to make sure that the Higgs field is efficiently tracking the minimum of its potential during the scanning process. This ensures that  the back-reaction from the exponential production of gauge bosons  is triggered when the VEV is at the electroweak scale. Hence, we impose that
\begin{equation}
\label{eq:tracking}
 \left| \frac{\dot{v}}{v^2}\right| \, \lesssim\, 1,
\end{equation}
where $v = (\Lambda^2 -g'\Lambda \phi)^{1/2}/\sqrt{\lambda}$ is the minimum of the Higgs potential given in \Eq{eq:relaxion_Lagrangian}.
The need for this conditions can be understood as follows. The mass of the $Z$ boson, and hence the time at which its tachyonic  production starts, depends on the value of the Higgs field $h$. After relaxation is over, $h$ will relax to the minimum $v$, so that $v$ controls the current value of the EW scale. Therefore, it is important that the evolution of $h$ and $v$ match, otherwise the relaxion field would stop as soon as $h$ satisfies \Eq{eq:tachyon}, while having a value of $v$ different from the current one.
\Eq{eq:tracking}  needs to be satisfied until the Higgs field value has reached the electroweak scale. 

Another necessary condition is that the average slow-roll velocity during the scanning has to be large enough to overcome the barriers generated by the cosine potential in \Eq{eq:relaxion_Lagrangian},
\begin{equation} \label{eq:phidotroll}
\dot{\phi}_{\textrm{roll}} \gtrsim \Lambda_b^2,
\end{equation}
where $\dot{\phi}_{\textrm{roll}} \sim V'_{\phi}/(3 H_I) + \delta(t)$ with $V'_{\phi}= g\,\Lambda^3$ and $\delta(t)$ being a  contribution due to the cosine potential.
At the same time, the average slow roll velocity should not exceed the cut-off, for the consistency of the effective theory: $V'_{\phi}/(3 H_I) \lesssim \Lambda^2$.
The sharp cut in the green region of Fig.(1) at $\Lambda\gtrsim 10^4 \GeV$ descends from this condition, after fixing $H_I$ to get the correct relic abundance (see below for more details).

Additionally, once the back-reaction has turned on, the barriers must be high enough to stop the relaxion evolution, requiring that
\begin{equation} 
\Lambda_b^4 \gtrsim g\, \Lambda^3 f'.
\end{equation}

We should also ensure that once relaxion is slowing down, the Higgs mass does not change  by an amount larger than the correct value, i.e.,
 \begin{equation} \label{eq:6H}
\Delta m_h \sim \frac{\Delta m_h^2}{m_h} \sim  \frac{1}{m_h} g' \Lambda \, \dot\phi \, \Delta t_\textrm{pp} \lesssim m_h\,.
\end{equation}

We impose that the kinetic energy lost by $\phi$ due to particle production is larger than the one gained by rolling down the potential,
\begin{equation}
\Delta K_\textrm{rolling}\, \lesssim\, \Delta K_\textrm{pp}.
\end{equation}
We estimate the two terms as $\Delta K_\textrm{pp}\sim \dot\phi^2 /2$ and $\Delta K_\textrm{rolling} \sim \frac{dK}{dt}  \Delta t_{\rm{pp}}$, where $dK/dt = -dV/dt \sim g \Lambda^3 \dot\phi$.  To be conservative [see \Eq{eq:phidotroll}], we take $\dot\phi^2/2 \sim\Lambda_b^4$.

On top of that, one should guarantee that the particle production is faster than the expansion rate,
\begin{equation}
\Delta t_\textrm{pp} < H_I^{-1},
\end{equation} 
so that the energy dissipation efficently slows down the relaxion field.

Furthermore, the scanning must have enough precision to resolve the electroweak scale. The mass parameter $\mu_h^2$ cannot vary more than the Higgs mass over one period of the  cosine potential,
\begin{equation} 
g' \Lambda\, \delta\phi =  g'\Lambda\, ( 2\pi f' )\lesssim m_h^2\,.
\end{equation}

In addition, it is crucial that the induced coupling to photons in \Eq{eq:relaxion coupling fermions photons} is suppressed enough, otherwise the dissipation from particle production would be relevant independently of the value of the Higgs mass. Then we have to impose that the produced photons are efficiently diluted by the cosmic expansion 
\begin{equation} \label{eq:dilution}
\Delta t_\gamma > H_I^{-1},
\end{equation}
where $\Delta t_\gamma \sim T^2 f_\gamma^3/\dot{\phi}^3$ with $f_\gamma$ given in \Eq{eq:f_gamma}. The relaxion induced coupling to photons through the Higgs mixing is very suppressed and the dilution requirement in \Eq{eq:dilution} for this contribution is trivially satisfied. 

A last condition concerns the restoration of the shift symmetry. After the relaxion has been trapped into one of  the wiggles, the temperature cannot be larger than the confinement scale, $T<\Lambda_b$, where, to be conservative, we assumed $\Lambda_b\approx\Lambda_c$ [see discussion below \Eq{eq:relaxion_Lagrangian}].  This condition is only relevant if the sector which generates the cosine potential gets in thermal equilibrium with the SM model. This can be estimated as follows. We assume that the barriers are generated by some confining gauge group, which is coupled to the relaxion via a term $(\phi/f') G'\widetilde G'$. Then, we naively estimate the rate for $g'g' \leftrightarrow ZZ$ interactions as $\Gamma\sim T^5 /(f^2 f'^2)$, which must be larger than the Hubble rate $H_I$ to achieve thermalization.

All in all, the conditions that apply to the parameters of our model are the following:


\begin{widetext}

\begin{align}
H_I \gtrsim \,& \frac{g\, \Lambda}{3} & \textrm{slow-roll velocity} \\
H_I \gtrsim\, & \frac{g' g\, \Lambda^4}{3 v_\ew^3 \lambda^{3/2}} & \textrm{Higgs  tracking the minimum} \\
H_I \lesssim\, & \frac{g\, \Lambda^3}{3 \Lambda_b^2} & \textrm{overcome the wiggles} \\
H_I \gtrsim & \left(\frac{10^{-4} g^5 \Lambda^{15}}{\sqrt{g_*} m_Z^3 \Lambda_b^8}\right)^{1/4}  & \textrm{efficient dissipation} \\
H_I \gtrsim & \left( \frac{10^{-4} g' g^4\, \Lambda^{13}}{\sqrt{g_*} m_h^2 m_Z^3 \Lambda_b^4}\right)^{1/4} & \textrm{small Higgs mass variation} \\
H_I \gtrsim\, & \textrm{Min}\left[\left(\frac{5}{3} \frac{g^2 \Lambda^6}{g_* \pi^2 \Lambda_b^4}\right)^{1/2}, \left(\frac{230\, m_Z^8\, g^2 \Lambda^6}{g_*^5 f'^8}\right)^{1/6}\right] & \textrm{no symmetry restoration} \\
H_I \gtrsim & \frac{\Lambda^2}{\MPl} & \textrm{inflaton potential dominates} \\
H_I \lesssim & \left(\frac{2\pi}{3}\right)^{1/3} g^{1/3}\Lambda & \textrm{classical rolling dominates} \\
H_I \gtrsim\, & \frac{16}{9 \pi^2 g^2_\ew} \frac{\dot{\phi}^3}{T^2 f_\gamma^3} & \textrm{photon dilution} \\
H_I \lesssim\, & \frac{16}{9 \pi^2 g^2_\ew}\frac{\dot{\phi}^3}{T^2 f^3} & \textrm{particle production fast} \\
g\lesssim\, & \frac{\Lambda_b^4}{\Lambda^3 f'} & \textrm{stopping condition} \\
g'\lesssim\, & \frac{m_h^2}{2\pi f'\Lambda} & \textrm{scanning with enough precision} \\
f' \gtrsim &\, \Lambda_b, \Lambda ~~ \textrm{and}~~ f \gtrsim \Lambda & \textrm{consistency of the EFT}
\end{align}

\end{widetext}

\begin{figure}
 \center
\includegraphics[width=.48\textwidth]{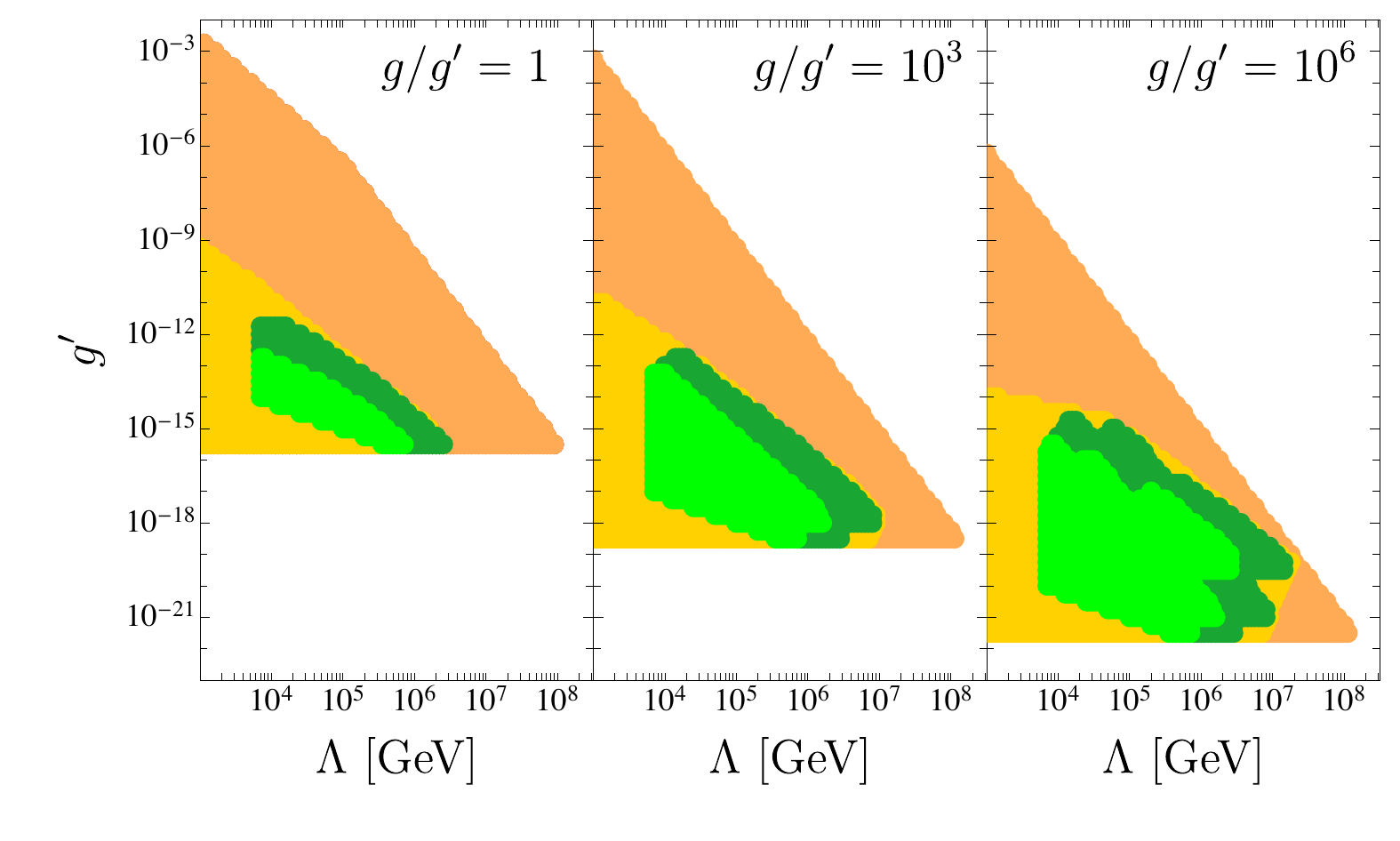}
\caption{Parameter space consistent with relaxation during inflation using particle production (pp) as the stopping mechanism.
Orange: $\phi$ is unstable; yellow: $\phi$ is overproduced; green: compatible with the relaxion being DM; light green: results after applying indirect detection bounds.}
\label{fig:gpLambda_colors}
\end{figure}

The coloured region in Fig.~\ref{fig:gpLambda_colors} shows the values of $\Lambda, g'$ for which the relaxion mechanism can be realized successfully, for a fixed ratio $g/g'$. To each point it corresponds a range in the other three free parameters.
 The colors correspond to different conditions which are imposed in order to make the relaxion a viable DM candidate, as we will detail in the next section.

\section{Relaxion as Dark Matter}\label{sec:DM}

The relaxion can be produced via vacuum misalignment and through thermal scattering. In the first case, after the relaxion gets stuck in one of the barriers,   it   will eventually start  to oscillate freely when particle production becomes inefficient, leading to a energy density which red-shifts as non-relativistic matter.  Since in our scenario the relaxation dynamics happens during inflation, the energy density stored in the field is diluted away and the misalignment contribution to the relaxion abundance is negligible  \cite{Hook:2016mqo} (see also Refs.~\cite{Espinosa:2015eda, Flacke:2016szy}).
The only  possibility   to produce a significant relaxion abundance  is then via scattering.

 A  population of relaxion particles is  produced through $a+b\leftrightarrow\phi + c$ interactions, where the species $a, b, \text{and}\, c$ belong to the SM and are in thermal equilibrium. The relaxion abundance  is controlled by the Boltzmann equation  \cite{Kolb:1990vq} 
\begin{equation} \label{eq:dYphi/dx}
\frac{d Y_\phi}{dx} = - \frac{\Gamma}{x H} \left(Y_\phi - Y_\phi^\eq\right), 
\end{equation}
where   $Y_\phi = n_\phi/s$ and $x=m_\phi/T$ with $n_\phi$ being the relaxion number density and $s$  the entropy density.
The equilibrium number density  of $\phi$ is $Y_\phi^\eq = n_\phi^\eq/s \approx 0.278/g_*$ where $g_*$ is the number of relativistic degrees of freedom, and $H$ is the Hubble rate. The quantity $\Gamma$ is given by the sum over the interaction rates,  $\Gamma \equiv \sum_i \Gamma_i $ with $\Gamma_i = n_{c_i} \langle \sigma v\rangle_i$, where the sum  includes gluon scattering~\cite{Masso:2002np}, Primakoff scattering (via $\phi \gamma \gamma$, $\phi Z \gamma$, and $\phi BB$, see, \eg, Refs.~\cite{PhysRevLett.59.2489, Cadamuro:2011fd, Flacke:2016szy}),  Compton scattering of leptons and quarks (via $\phi \bar{l}l$ and $\phi \bar{q}q$,  see, \eg, Refs.~\cite{PhysRevLett.59.2489, Flacke:2016szy}), and Primakoff and Compton processes through the mixing with the Higgs (see, \eg, Ref.~\cite{Flacke:2016szy}).
The pion-conversion processes $\pi^0 N \rightarrow \phi\, N$ and $\pi^0 \pi^0 \rightarrow \phi\, \pi^0$ can be neglected, as they are only active for a short time around the QCD phase transition, and their integrated rate is thus negligible. Assuming that the initial $\phi$  density  is negligible,  $Y_\phi(x_0)=0$, the solution of \Eq{eq:dYphi/dx} is 
\begin{equation} \label{eq:Yphi}
Y_\phi(x) = Y_\phi^\eq \left[1 - \textrm{exp}\left(-\int_{x_0}^x \frac{\Gamma}{x' H} dx'\right)\right] \,,
\end{equation}
where $T_0 = m_\phi / x_0$ can be identified with the reheating temperature (we will comment more on this below).
If the  integrand is large, then $Y_\phi(x) \approx Y_\phi^\eq$ and the correct DM abundance can only be met for a very light (thus hot) DM component.
We must therefore be in the opposite situation, in which  the integrand  is  smaller than one, and we can approximate $Y_\phi$ by
\begin{equation} \label{eq:Yphiapprox}
Y_\phi(x) \approx Y_\phi^\eq  \int_{x_0}^x \frac{\Gamma}{x' H} dx'.
\end{equation}
All the processes listed above are suppressed by the same physical scale $f$. The corresponding terms induced by the relaxion-Higgs mixing give a negligible contribution to the final $\phi$ abundance. 
The dominant production channels are Compton scattering off fermions $\gamma + F \leftrightarrow \phi + F$ and, above the QCD scale, the corresponding gluonic process $g + q \leftrightarrow \phi + q$, both mediated by the axion-fermion coupling of \Eq{eq:f_fermions}. At any given temperature, the leading process involves the most massive fermion which is relativistic at that time, due to the fact that the interaction rate scales as $m_F^2$.
Using the expressions for the interaction rates $\Gamma_i$ in Refs.~\cite{Masso:2002np, PhysRevLett.59.2489, Cadamuro:2011fd, Flacke:2016szy}, we checked explicitly that all the other processes give a subdominant contribution to the relic abundance. We can also neglect all the interference terms, since at each temperature the subdominant processes are highly suppressed compared to the main one.
The interaction rates for a fermion $F$ are given by
\begin{align}
\Gamma_{C,\gamma}  &  \approx \frac{3\zeta(3)}{\pi^2}\alpha_\mathrm{em} \frac{m_F^2 T}{f_F^2} \,, \\
\Gamma_{C,g} &  \approx \frac{36\zeta(3)}{\pi^2}\alpha_s \frac{m_F^2 T}{f_F^2} \,, 
\end{align}
where $f_F$ is given in \Eq{eq:f_fermions}.
In order to guarantee that the interactions are out-of-equilibrium ($\Gamma_i/H < 1$) and that the relaxion never enters in thermal equilibrium with the SM bath, we will need a rather low $T_0$.

The relaxion decays through the loop-induced couplings to photons and SM fermions as in \Eq{eq:relaxion coupling fermions photons},
by  the leading interaction  with the electroweak gauge bosons in \Eq{eq:relaxion_Lagrangian}, and via the mixing with the Higgs [for which we used the results in Ref.~\cite{Bezrukov:2009yw} multiplied by $\theta^2$ of \Eq{eq:mixing}].
We only consider 2-body decays in our analysis.

As we shall see,  relaxion dark matter is in the keV range.   In this mass ballpark the relaxion can only decay into photons and  neutrinos. The decay into photons proceeds through the mixing with the Higgs and through the loop-induced coupling of \Eq{eq:relaxion coupling fermions photons}. For simplicity, we  assume  that neutrinos are Majorana fermions, in which case the decay in this channel is suppressed compared to the one into photons as it proceeds via higher dimensional operators (see, \eg, Ref.~\cite{Bauer:2017ris}). If neutrinos are Dirac fermions, this can be the dominant decay channel. Nevertheless, the bounds from indirect detection on the DM decaying into photons (see next section)  imply  stronger constraints on the relaxion lifetime.

\medskip

\section{Results and discussion}\label{sec:results}

We performed a  scan looking for points in $\{\Lambda,g',\Lambda_b,f'\}$ which can satisfy the DM hypothesis. For each point, the value of $H_I$ is fixed by plugging \Eq{eq:f} into \Eq{eq:Yphiapprox}, and then requiring to match the observed DM abundance. This value has to be compared with the range allowed from the conditions on particle production.
The result is shown in Fig.~\ref{fig:gpLambda_colors}. The green region is the one where the relaxion is stable,  all the bounds  on a successful relaxation with particle production  are simultaneously satisfied and the relaxion abundance matches the observed DM one (for a given range of $\Lambda_b$ and $f'$). The light green part is the one in which, additionally, the constraints from indirect detection are satisfied.
In the yellow region the relaxion can be   stable, but it is overproduced. Finally, in the orange region the relaxion's lifetime is shorter than the age of the universe.
Table \ref{tab:range} shows the allowed parameter space for  three benchmarks.

On top of the above constraints, we applied a lower bound on the DM mass from structure formation. The free-streaming length $\lambda_{\textrm{NR}}$ is constrained by observations from Lyman-$\alpha$ forest \cite{Viel:2013apy,Irsic:2017ixq}, which are in tension with a thermal relic below a few keV. We  estimate the free-streaming length as \cite{Kolb:1990vq}:
\begin{align} \nonumber 
\lambda_{\textrm{FS}}  &= a_0 \int_{t_\textrm{FI}}^{t_{\textrm{NR}}} \frac{1}{a(t')}dt' + a_0 \int_{t_{\textrm{NR}}}^{t_{\textrm{EQ}}} \frac{a(t_{\textrm{NR}})}{a^2(t')}dt'\\ 
 &=  \frac{a_0}{a(t_{\textrm{NR}})}    t_{\textrm{NR}} \left[2 - 2\left(\frac{t_{\textrm{FI}}}{t_{\textrm{NR}}} \right)^{1/2} + \ln\left(\frac{t_{\textrm{EQ}}}{t_{\textrm{NR}}}\right)\right], 
\end{align}
where $a(t)$ is the scale factor as a function of time within $a_0$ being the scale factor today, $t_{\textrm{FI}}$ refers to the time when  DM abundance freezes-in,   after which  it free-streams relativistically until  $t_{\rm{NR}} \gg t_{\textrm{FI}}$ when it becomes  non-relativistic. The relaxion then free-streams non-relativistically up to the matter-radiation equality at $t_{\rm{EQ}}$. The time $t_{\textrm{NR}}$ can be easily obtained if the DM velocity distribution is thermal. However,  in our scenario the DM  distribution function can depart from a thermal distribution as the relaxion is produced out-of-equilibrium, which may weaken these bounds (see Refs.~\cite{Bernal:2017kxu, Heeck:2017xbu} and references therein). It would be important  to further explore such feature which we leave for future work. Here we simply impose that the relaxion should be heavier than 2 keV.

\begin{figure}
\center
\includegraphics[width=.48\textwidth]{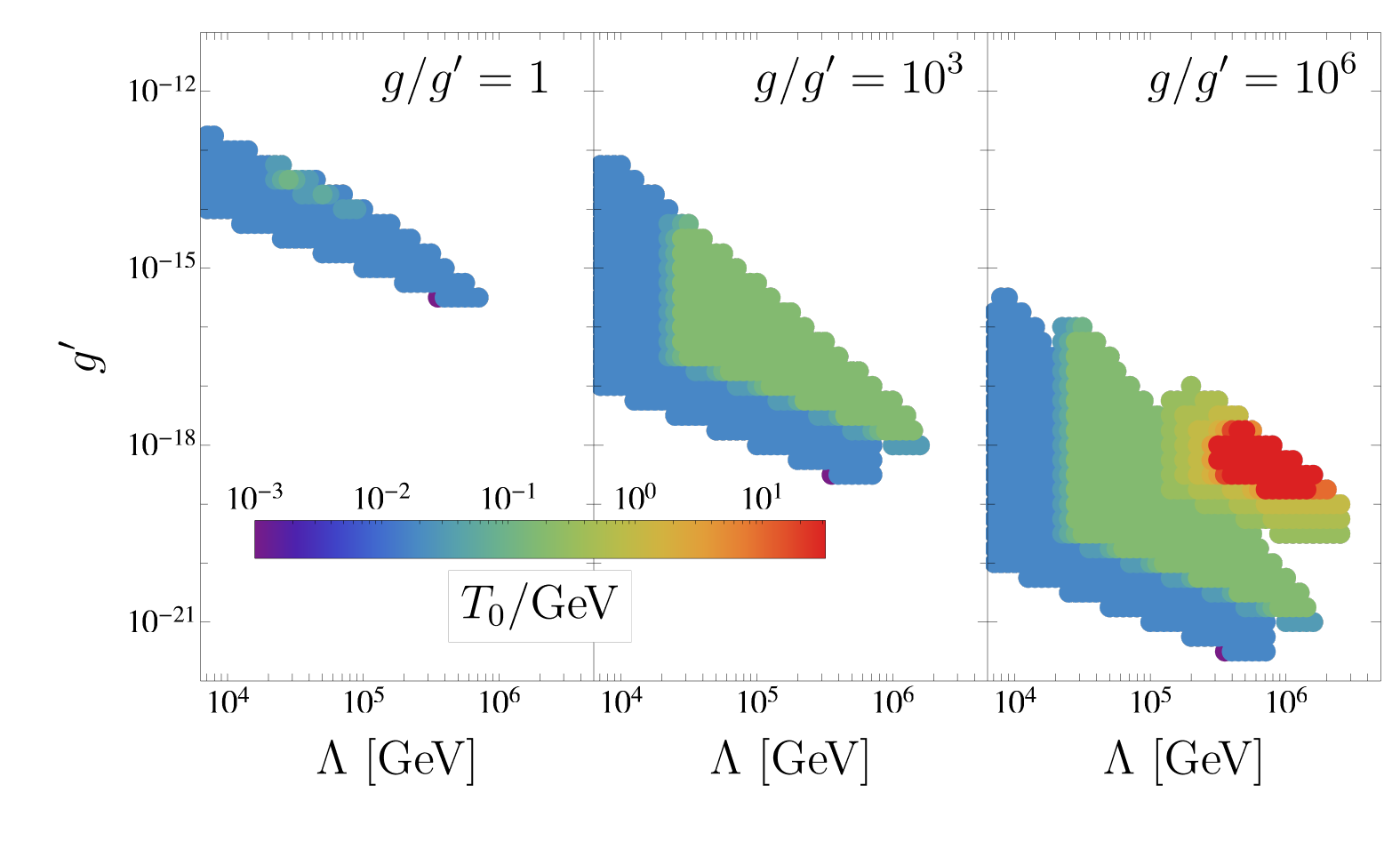}
\caption{Allowed dark matter region as a function of the reheating temperature. The region shrinks for higher temperature $T_0$.}
\label{fig: parameter space TRH}
\end{figure}

Figure~\ref{fig: parameter space TRH} shows how the allowed region depends on the value of the reheating temperature.  A drawback of our scenario is that the reheating temperature is rather low, $T_0 \lesssim 100 \MeV, 200\MeV, 30\GeV$ for $g/g' = 1, 10^3, 10^6$, to avoid overabundance. Ultimately, this is due to the UV sensitivity of the production mechanism. If thermal equilibrium is reached, in this mass range the relaxion would be overabundant by a factor of $10-100$.
Oppositely, the correct relic abundance could only be obtained for a correspondingly lighter particle, which would then be too light to comply with the warm DM mass lower bound.

Measurements of the abundance of light elements,  large scale structure data, and anisotropies of cosmic microwave background  temperature constrain late-time entropy production, which then  restricts  the reheating temperature to be larger than $1-4\MeV$~\cite{Kawasaki:1999na, Kawasaki:2000en, Hannestad:2004px, Ichikawa:2005vw, DeBernardis:2008zz, deSalas:2015glj}.
Below this temperature, the universe behaves like radiation, and only very small entropy injections are possible.

A low reheating temperature can be achieved even if the temperature of the SM plasma at the end of inflation rises to much higher values, during a phase of entropy injection~\cite{Turner:1983he, Scherrer:1984fd, Giudice:2000ex}. This is indeed expected to happen when the inflaton decays perturbatively into SM particles. 
The temperature first rises to a maximal value $T_\mathrm{max} \sim T_0^{1/2} (H_I \MPl)^{1/4}$, then it decreases with the typical dependence on the scale factor $T\propto a^{-3/8}$. This behaviour proceeds until the decay of the inflaton  ceases at a time of order the inverse decay width of the inflaton, at which radiation dominance begins with the standard $T\propto a^{-1}$ behaviour. During the reheating phase, entropy is continuously created, and the Hubble rate scales as $H\sim[g_*(T)/g_*^{1/2}(T_0)]T^4/(T_0^2\MPl)$ (expansion is faster for lower reheating temperature).  In such a scenario, the abundance of relic particles are altered compared to the standard radiation dominance calculation. On the one hand, particles with mass larger than the reheating temperature can be copiously produced~\cite{Chung:1998rq}. On the other hand, which is the case relevant here, particles with a freeze-out temperature larger than $T_0$ are diluted by entropy injection, and their abundance is smaller than in the standard freeze-out computation. As an example, in Refs.~\cite{Giudice:2000dp, Giudice:2000ex}  was argued that SM model neutrinos in the keV mass range (hence now excluded) could have the right relic abundance to be a warm DM candidate.
The relic abundance of long-lived particles in a low reheating scenario is studied in the literature for many kind of DM candidates, such as sterile neutrinos~\cite{Gelmini:2004ah, Yaguna:2007wi, Gelmini:2008fq}, supersymmetric particles and more generic WIMPs~\cite{McDonald:1989jd, Giudice:2000ex, Fornengo:2002db, Gelmini:2006pw, Drees:2006vh, Gelmini:2006pq}, heavy particles~\cite{Chung:1998rq}, axions~\cite{Dine:1982ah, Steinhardt:1983ia, Yamamoto:1985mb, Lazarides:1987zf, Kawasaki:1995vt, Giudice:2000ex, Grin:2007yg, Visinelli:2009kt}.
Here we just assume that at $T_0$ the relaxion abundance is negligible, and that its relic abundance is built up during radiation dominance.

Finally, let us mention that baryogenesis mechanisms that require a large temperature are also viable in this scenario. As an example, electroweak baryogenesis is viable in this case for $T_0$ as small as 1~GeV~\cite{Davidson:2000dw, Giudice:2000ex}, thus favoring large values of the ratio $g/g'$.
While this is an interesting option, a concrete realization which connects the relaxion to DM and to baryogenesis  is beyond the scope of the present work.

Strong constraints on the model come from the observations of the galactic and extra-galactic diffuse X-ray and $\gamma$-ray background.  We consider the constraints on decaying DM from Ref.~\cite{Essig:2013goa} which uses the diffuse photon spectra data from different satellites.  For our parameter space, which comprises masses around the $\keV$ range, the relevant bounds are given by the satellites HEAO-1 \cite{Gruber:1999yr} and INTEGRAL \cite{Bouchet:2008rp}.  In \Fig{fig:gpLambda_colors} we show in  light green  the region  in agreement with the bounds  on the lifetime of a scalar DM  decaying into two photons, $\tau_\phi\gtrsim 10^{26-28} \s$ for $m_\phi > 5\keV$ \cite{Essig:2013goa}. This constrains the relaxion mass   to be $m_\phi \lesssim \{5\keV, 5\keV, 17 \keV \}$, respectively for the three benchmarks in Tab.~\ref{tab:range}. Extrapolating the bound from Ref.~\cite{Essig:2013goa} to lower masses further constrains the parameter space, but the results are qualitatively similar.
This places  relaxion DM in a  knife-edge  position: on the one hand, new results from indirect searches in the keV mass range could rule out this scenario; on the other hand, a numerical solution of the Boltzmann equation could weaken the lower bound on the relaxion mass, thus opening the parameter space for lighter DM.

\begin{table}
\renewcommand{\arraystretch}{2}
\begin{tabular}{cccc}
\hline\hline
& $g/g' = 1$ & $g/g' = 10^3$ & $g/g' = 10^6$ \\ \hline

\scriptsize  $ \displaystyle \frac{\Lambda}{\mathrm{GeV}}$ & \scriptsize $ 7\cdot 10^3 - 7\cdot 10^5$ & \scriptsize $7\cdot 10^3  - 2\cdot 10^6$ & \scriptsize $7\cdot 10^3  - 3\cdot 10^6$ \\

\scriptsize $  \displaystyle g'$ & \scriptsize $  3\cdot 10^{-16} - 2\cdot 10^{-13}$ & \scriptsize $ 3\cdot 10^{-19} - 6\cdot 10^{-14}$ & \scriptsize $ 3\cdot 10^{-22} - 3\cdot 10^{-16}$ \\

\scriptsize $\displaystyle \frac{m_\phi}{\mathrm{keV}}$ & \scriptsize $2-4$ & \scriptsize $2-4$ & \scriptsize $2 - 17$ \\

\scriptsize $ \displaystyle \frac{f'}{\mathrm{GeV}}$ & \scriptsize $ 3\cdot 10^{10}- 2\cdot 10^{13}$ & \scriptsize $3\cdot 10^{10}-6\cdot 10^{14}$ & \scriptsize $3\cdot 10^{10}-3\cdot 10^{16}$ \\

\scriptsize $ \displaystyle \frac{\Lambda_b}{\mathrm{GeV}}$ & \scriptsize $3\cdot 10^2 - 8\cdot 10^3 $ & \scriptsize $3\cdot 10^2 - 4\cdot 10^4 $ & \scriptsize $3\cdot 10^2 - 4 \cdot 10^5 $ \\

\scriptsize $\displaystyle \frac{f}{\mathrm{GeV}}$ & \scriptsize $3\cdot 10^5 - 4\cdot 10^6$ & \scriptsize $3\cdot 10^5 - 9\cdot 10^6$ & \scriptsize $3\cdot 10^5 - 2\cdot 10^9$ \\

\scriptsize $\displaystyle \frac{H_I}{\mathrm{GeV}}$ & \scriptsize $2\cdot 10^{-11} - 3\times 10^{-7}$ & \scriptsize $2\cdot 10^{-11} - 10^{-6}$ & \scriptsize $2\cdot 10^{-11} - 4\cdot 10^{-6}$ \\

\scriptsize $\displaystyle \frac{T_0}{\mathrm{GeV}}$ & \scriptsize $10^{-3} - 10^{-1}$ & \scriptsize $10^{-3} - 2\cdot 10^{-1}$ & \scriptsize $10^{-3} - 3\cdot 10^1$ \\

\scriptsize $  \displaystyle N_e$ & \scriptsize $0.3 - 5\cdot 10^6$ & \scriptsize $3\cdot 10^{2} - 5\cdot 10^{9}$ & \scriptsize $3\cdot 10^{5} - 5\times 10^{12}$\\

\scriptsize $  \displaystyle \theta$ & \scriptsize $10^{-12} - 2\times 10^{-11}$ & \scriptsize $10^{-15} - 9\cdot 10^{-12}$ & \scriptsize $10^{-18} - 5 \cdot 10^{-14}$\\

\hline\hline

\end{tabular}
\caption{\label{tab:range} Allowed parameter space after taking into account the bounds from indirect DM searches, for the benchmarks $g/g' = 1, 10^3$ and $10^6$. The last two lines show the minimal number of e-folds that inflation should last to allow for relaxation to complete, and the value of the relaxion-Higgs mixing angle.}
\end{table}

Other important constraints are given by  astrophysical probes \cite{Craig:2018kne}.  The relaxion coupling to electrons  (see \Eq{eq:f_fermions})  is  constrained from red giants observations, which results in a lower bound for the coupling in \Eq{eq:relaxion_Lagrangian} of $f\gtrsim 3\times 10^7 \GeV$  for  $m_\phi \lesssim 10^{-5}\GeV$. Even more stringent is the bound from Supernova 1987A, which for  $m_\phi \lesssim 0.1 \GeV$, disfavors $f\lesssim 10^8 \GeV$.  It should be noticed that the uncertainties associated with the bounds derived from astrophysical sources are typically within an order of magnitude \cite{Craig:2018kne, Krnjaic:2015mbs, Flacke:2016szy}.  This implies that part of our parameter space in Tab.\,\ref{tab:range} is in tension with such bounds. On the other hand, the parameter space for  successful relaxation with particle production is also subject to some variation.
The requirements for a successful  particle production mechanism could be relaxed by considering the relaxion velocity. This has two natural reference values: the first is the slow-roll velocity $\dot\phi_\mathrm{roll}\sim V_\phi'/(3 H_I)$, while the second is the minimal velocity to overcome the wiggles $\dot\phi_b \sim \Lambda_b^2$. The two are related by $\dot\phi_\mathrm{roll}>\dot\phi_b$. In deriving the relations, we always chose the value that lead to the most conservative bound. Some of the conditions on particle production could therefore be weakened by choosing a different value for the velocity, but we do not pursue this possibility further.

\section{Conclusions}\label{sec:conclusions}

 In this work, we showed that the relaxion mechanism can naturally provide a phenomenologically viable warm DM candidate in the keV mass range. We identified the relevant parameter space in the scenario in which relaxation happens during inflation, using particle production as a source of friction. We discussed astrophysical and indirect detection constraints on the model.

Recently, there has been an increasing interest in DM direct detection experiments that can probe the sub-MeV  mass range (see, \eg, Refs.~\cite{Hochberg:2015pha, Schutz:2016tid}).
The relaxion would be a well motivated DM candidate in the $\keV$ range, which  encourages new studies in this mass ballpark.

It would be interesting to further explore the consequences of such a model on structure formation, and perform a dedicated analysis of the indirect detection bounds. We leave these studies for future work.

\section*{Acknowledgments}

We are happy to thank  Géraldine Servant for encouragement, discussions and helpful comments. We are also grateful to  Filippo Sala for valuable comments. NF  thanks the organizers and participants of the CERN TH Institute on Physics at the LHC and Beyond for interesting discussions while part of this work was completed.

\bibliography{PPRelDMBib}

\end{document}